\documentclass[preprint,amsmath,amssymb,nofootinbib,superscriptaddress]{revtex4}

\usepackage{amsmath,graphicx,color,epsfig}

\begin{document}


\title{Towards the continuum coupling in nuclear lattice effective field theory I: A three-particle model}

\author{J.-J. Wu\footnote{wujiajun@ucas.ac.cn}}
\affiliation{School of Physical Sciences, University of Chinese Academy of Sciences, Beijing
  100049, China}

\author{Ulf-G. Mei{\ss}ner\footnote{meissner@hiskp.uni-bonn.de}}
\affiliation{Helmholtz-Institut f\"ur Strahlen- und Kernphysik and Bethe Center for
  Theoretical Physics, Universit\"at Bonn, D-53115 Bonn, Germany}
\affiliation{Institute for Advanced Simulation, Institut f\"ur Kernphysik,
  and J\"ulich Center for Hadron Physics, Forschungszentrum J\"ulich, D-52425 J\"ulich, Germany}
\affiliation{Tbilisi State University, 0186 Tbilisi, Georgia}

\begin{abstract}
  Weakly bound states often occur in nuclear physics. To precisely understand their properties,
  the coupling to the continuum should be worked out explicitely. In a first step, we
  use a simple nuclear model in the continuum and on a lattice to investigate the influence
  of a third particle  on a loosely bound state of a particle and a heavy core.
\end{abstract}

\maketitle



\section{Introduction}

Along the nuclear chart, there are a number of weakly bound states like in case of halo nuclei
or for isotopes close to the drip lines. These states are characterized by binding energies
in the keV range rather than the few MeV typical for nuclear binding. Such loosely
(or weakly) bound states are thus closely located to decay thresholds and the corresponding
continuum of states. Under such circumstances the coupling of such a bound state
to the continuum can no longer be neglected, for reviews see
e.g.~\cite{Dobaczewski:2007jc,Michel:2008pt,Meng:2015hta}.
For conventional nuclear models, like e.g. the shell model or the no-core-shell model, the coupling to the
continuum based e.g.  on Berggren's representation~\cite{Berggren:1968zz,Berggren:1993zz}, that
treats bound, resonant and continuum states on the same footing, is well established,
see e.g.~\cite{Grasso:2001hf,Papadimitriou:2013ix,Sun:2017unq}. In addition, {\em ab initio} calculation
for systems such as $^4$He$+n+n$ and $A=7$ isotopes have been performed including continuum
effects~\cite{Baroni:2013fe,Romero-Redondo:2014fya,Vorabbi:2019imi}.

Nuclear lattice effective field theory (NLEFT) is a novel method for performing {\em ab initio} calculations
in nuclear structure and reaction physics~\cite{Lee:2008fa,Lahde:2019npb}. The basic idea is to discretize
space-time on a finite volume $L^3\times L_t$, with $L \, (L_t)$ the spatial (temporal) size. Nucleons are
placed on the lattice sites and their interaction are given in terms of properly modified chiral potentials,
consisting of pion exchanges and short-distance operators. Strong isospin-breaking effects and the
long-ranged Coulomb potential are also included, leading to a number of intriguing results, like e.g.
the {\em ab initio} calculation of the Hoyle state in $^{12}$C \cite{Epelbaum:2011md}
or the first microscopic calculation of low-energy $\alpha$-$\alpha$ scattering \cite{Elhatisari:2015iga}.
What is missing in this framework is the coupling to the continuum. Of course, on the lattice we have
only real-valued energies, so a direct application of the Berggren approach is not possible. However,
as shown by L\"uscher in this seminal work, the complex-valued scattering phase shift can be mapped
onto the volume-dependence of the lattice energy levels~ \cite{Luscher:1986pf,Luscher:1990ux}.
We seek a similar formalism to explicitely describe the continuum coupling.

In this work, we use a simple model of a heavy core $A$ coupled to one or two nucleons $N$,
as described in Sect.~\ref{sec:model}. In Sect.~\ref{sec:2body} the consider $AN\to AN$
scattering and adjust the $AN$ interaction such that a very weakly bound state emerges. Using the
Hamiltonian formalism of
Refs.~\cite{Hall:2013qba,Wu:2014vma,Liu:2015ktc,Wu:2016ixr,Liu:2016wxq,Wu:2017qve,Li:2019qvh}
we calculate the energy levels of this
system in a finite volume. The full $ANN$ system is considered in Sect.~\ref{sec:3body}, where we
adjust the parameters so that there is no three-particle bound state and we can thus study the
influence of the unbound third particle on the $AN$ scattering matrix. We conclude with a summary and
outlook in Sect.~\ref{sec:sum}. The Appendix contains a short discussion of the normalization
of the scattering equation used.


\section{The model}
\label{sec:model}

Consider a three-particle model ($ANN$ system), with the mass of the $A$ particle about 10 times
the mass of the $N$ particle with mass $m$. The $A$ particle thus mimics the nuclear core.
To be specific, let us calculate $AN\to AN$ scattering. For simplicity, we use a separable
potential of the form
\begin{equation}
V^{AN}_H(p, p') = \frac{1}{\sqrt{2\omega_{A}(p)2\omega_{N}(p)}}
g f(p,\Lambda)f(p',\Lambda)\frac{1}{\sqrt{2\omega_{A}(p)2\omega_{N}(p)}}~,
\label{eq:VAN}
\end{equation}
with the regulator function
\begin{equation}
f(p,\Lambda)=\frac{\Lambda^2}{p^2+\Lambda^2}~,
\label{eq:f}
\end{equation}
with $\omega_i(q)=\sqrt{m_i^2+q^2}$
and $g$ is the  coupling constant. The normalization is explained in the Appendix.
In what follows, we set $m_A = 10\,$GeV, $m = \Lambda =1\,$GeV.
We are interested in the case that the $AN$ system has a very weakly bound state $B$
with keV binding energy, so the coupling $g$ will be tuned accordingly. 

Then we can construct the Hamiltonian in the finite volume and find its eigenvalues~\cite{Wu:2014vma}. 
The Hamiltonian matrix is defined as follows:
\begin{eqnarray}
H&=&H_0+H_I,\nonumber\\
\left(H_{0}\right)_{ij}&=&\delta_{ij}\left(\omega_A(k_i)+\omega_N(k_i)\right),\\
\left(H_I\right)_{ij}&=&
\frac{\sqrt{C_3(i)C_3(j)}}{4\pi}
\left(\frac{2\pi}{L}\right)^3
V_H(E,k_i,k_j),
\end{eqnarray} 
where  $k_i=\sqrt{i}2\pi/L$ and $C_3(i)$ represents the number of ways to sum the square of
the three integers to equal $i$. Further, the factor $(\sqrt{C_3(i)C_3(j)})/(4\pi)
(2\pi/L)^3$ is due to the quantization in a finite box with size $L$ as explained in
Refs.~\cite{Hall:2013qba, Wu:2014vma}.

\begin{figure}[t!]
\centering
\includegraphics[width=0.86\textwidth]{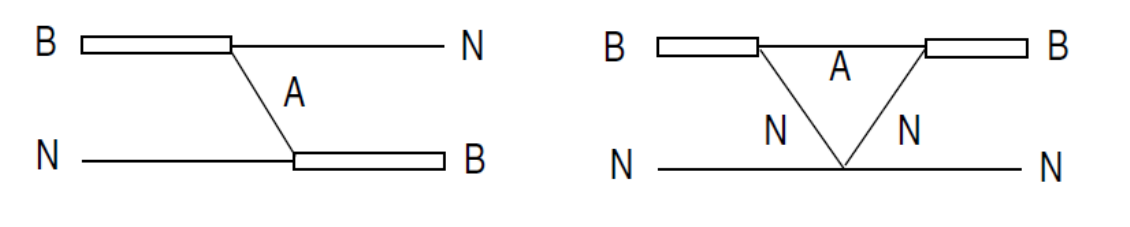}
\caption{Effective diagrams for the $BN \to BN$ process}
\label{fig:BNdiag}
\end{figure} 

For the full $ANN$ system with a fixed total momentum, we have two free momenta. This will lead to
a  Hamiltonian matrix in the finite volume with a huge dimension. For simplification, we thus
consider the  $BN$ system instead of $ANN$, that means we use a version of the dimer approximation,
see e.g. \cite{Kaplan:1996nv}, reminiscent of the so-called Faddeev fixed center approximation,
see e.g.~Refs.~\cite{Kamalov:2000iy,Zhang:2019ykd}.
We thus consider the scattering process $BN\to BN$.

The left diagram of Fig.~\ref{fig:BNdiag} shows the attractive interaction between $B$ and
$N$ since the $AN$  system has a weak attractive interaction.
To calculate this diagram, we need to get the coupling of $B\to AN$ process.
Since the $B$ is a loosely bound state of $AN$, one gets the coupling  from the amplitude of $AN \to AN$
around the pole position of $B$ as follows,
\begin{eqnarray}
T^{AN}_H(E\sim m_B, q=q_0(E), q'=q_0(E))&=&\frac{1}{2m_{A}}\frac{1}{2m_{N}}4\pi
\frac{\tilde{g}^2}{2m_B(E-m_B)},
\end{eqnarray} 
where $q_0(E)$ is the on-shell momentum with energy $E$, and $T^{AN}_H$ is obtained from
Eq.(\ref{eq:Hscattering})
with the potential $V^{AN}_H$.
The factor $1/(2m_{A}) \cdot 1/(2m_{N})$ originates from the the difference between $V_H$ and $V_L$
(see the Appendix), the momentum is on-shell, so it is close to the mass of the particle, and the
factor $4\pi$ is from  the angular integration, since we only consider the s-wave.
Further, the coupling  $\tilde{g}$ has dimension energy.
With that, the potential of $BN \to BN$ from $A$-exchange takes the form
\begin{eqnarray}
V^{BN1}_H(p, p')&=&
\frac{2\pi}{\sqrt{2\omega_B(p)2\omega_N(p)}}
\nonumber\\
&\times& \int  d \cos\theta \frac{\tilde{g}^2}
{(E_A^2)-(\vec{p}-\vec{p}\,')^2-m^2_A}
\frac{1}{\sqrt{2\omega_B(p')2\omega_N(p')}},\label{eq:VBN}\\
E_A^2&=&\frac{1}{2}\left((\omega_B(p)-\omega_N(p'))^2+(\omega_N(p)-\omega_B(p'))^2\right)~.
\end{eqnarray} 
Since our potential should be independent of the total energy, we take the average of the two processes
$B\to AN$ and $NA\to B$. Next,  we need to pick out the s-wave contribution of this diagram,
so we perform  the angular integration  between $\vec{p}$ and $\vec{p}\,'$.
At last, the equation for the potential takes the form
\begin{eqnarray}
V^{BN1}_H(p, p')&=&
\frac{\tilde{g}^2}{\sqrt{2\omega_B(p)2\omega_N(p)2\omega_B(p')2\omega_N(p')}}
\frac{\pi}{pp'}\nonumber\\
&\times&\ln\left(\frac{m_B^2+m_N^2-m_A^2-(\omega_B(p)\omega_N(p')+\omega_N(p)\omega_B(p'))+2pp'}
{m_B^2+m_N^2-m_A^2-(\omega_B(p)\omega_N(p')+\omega_N(p)\omega_B(p'))-2pp'}\right)~.\nonumber\\
\end{eqnarray} 
Note that  this potential should be negative, because in Eq.~(\ref{eq:VBN}) the propagator
of the exchanged $A$ particle is negative.

Now let us consider the  contribution from the right diagram of Fig.~\ref{fig:BNdiag}.
This includes a triangle loop, and the main interaction is the $NN \to NN$ interaction.
First,  the $NN \to NN$ interaction can be  written as,
\begin{eqnarray}
V^{NN}_H(p, p') &=& \frac{1}{\sqrt{2\omega_{N}(p)2\omega_{N}(p)}}
g_{NN} f(p,\Lambda)f(p',\Lambda)\frac{1}{\sqrt{2\omega_{A}(p)2\omega_{N}(p)}}~,
\label{eq:VNN}
\end{eqnarray}
where the regulator function is chosen the same as for the $AN$ interaction.
In this model, we want to describe the situation that  $BN$ system can not form a bound state, therefore
the coupling $g_{NN}$ is only parameter which allows to achieve that.

Next, we work out the potential based on the diagram on the right side of Fig.~\ref{fig:BNdiag}:
\begin{eqnarray}
V^{BN2}_H(p, p') &=& \!\!\frac{2\pi}{\sqrt{2\omega_{B}(p)2\omega_{N}(p)}}
\int d\cos\theta\, V^{BN2}_L(\vec{p},\vec{p}')
\frac{1}{\sqrt{2\omega_{B}(p')2\omega_{N}(p')}},\nonumber\\
\label{eq:VBN2}
\end{eqnarray}
where
\begin{eqnarray}
V^{BN2}_L(\vec{p},\vec{p}') &=& 
\int d^4q ~\tilde{g}^2 
\frac{1}{q^2_0-\vec{q}^2-m^2_A}
\frac{1}{(\omega_B(p)-q_0)^2-(\vec{p}-\vec{q})^2-m^2_N)}\nonumber\\
&\times& \frac{1}{(\omega_B(p')-q_0)^2-(\vec{p}'-\vec{q})^2-m^2_N}
T^{L}_{NN}
\label{eq:VBN2L}
\nonumber \\
&=&
\int d^3\vec{q} ~\tilde{g}^2 
\frac{1}{2\omega_A(q)}
\frac{1}{(\omega_B(p)-\omega_A(q))^2-(\vec{p}-\vec{q})^2-m^2_N}\nonumber\\
&\times& \frac{1}{(\omega_B(p')-\omega_A(q))^2-(\vec{p}'-\vec{q})^2-m^2_N)}
T^{NN}_{L}~,
\end{eqnarray}
where $T_{L}^{NN}$ is the amplitude of $NN \to NN$.
In the calculation of  $T_{L}^{NN}$, we make some further assumptions.
First, we assume $T_{L}^{NN} \sim V_{L}^{NN}$, which should be acceptable as we are not interested
in the detailed structure of the $NN$ scattering amplitude. Also, we require this interaction in
a boosted frame.  Although the form of $V_L$ is not Lorentz invariant, we can rewrite the potential
in a special form and define all inputs the center-of-mass (CM) system.
This means that we write $V^{NN}_L$ as    
\begin{eqnarray}
T^{NN}_{L} &\sim& V^{NN}_{L}=g_{NN} f(k^*,\Lambda)f(k'^*,\Lambda) 
,\label{eq:VBN2L}\\
k^{*\,2} &=&E_{NN}^2/4-m_N^2, \\
E_{NN}^2&=&\left(\sqrt{(\vec{q}-\vec{p})^2+m_N^2}+\sqrt{\vec{p}^2+m_N^2}\right)^2-q^2,\\
k^{'*\,2}&=&E_{NN}^{'2}/4-m_N^2, \\
E_{NN}^{'2}&=&\left(\sqrt{(\vec{q}-\vec{p}^{\,'})^2+m_N^2}+\sqrt{\vec{p}^{\,'2}+m_N^2}\right)^2-q^2.\label{eq:ENN'}
\end{eqnarray}
Then at last we can get $V_H^{BN2}$ as defined in Eq.(\ref{eq:VBN2}-\ref{eq:ENN'})
 as follows:
\begin{eqnarray}
V^{BN2}_H(p, p') &=& \!\!\frac{4\pi^2}{\sqrt{2\omega_{B}(p)2\omega_{N}(p)2\omega_{B}(p')2\omega_{N}(p')}}
\int q^2dq \frac{\tilde{g}^2 g_{NN}}{2\omega_A(q)} H(p,q)H(p',q),
\label{eq:VBN2cal}
\end{eqnarray}
where
\begin{eqnarray}
 H(p,q) &=& \int d\cos\theta \frac{1}{m^2_B+m^2_A-m^2_N-2\omega_B(p)\omega_A(q)+2pq\cos\theta} 
\nonumber \\
&\times&\frac{4\Lambda^2}{\left(\sqrt{q^2+p^2-2pq\cos\theta+m_N^2}+\sqrt{p^2+m_N^2}\right)^2-q^2-4m_N^2+4\Lambda^2}~,
\label{eq:Hpq}
\end{eqnarray}
which can easily be evaluated numerically.

\section{Results for $2\to2$ scattering}
\label{sec:2body}

\begin{figure}[t!]
\centering
\includegraphics[width=0.45\textwidth]{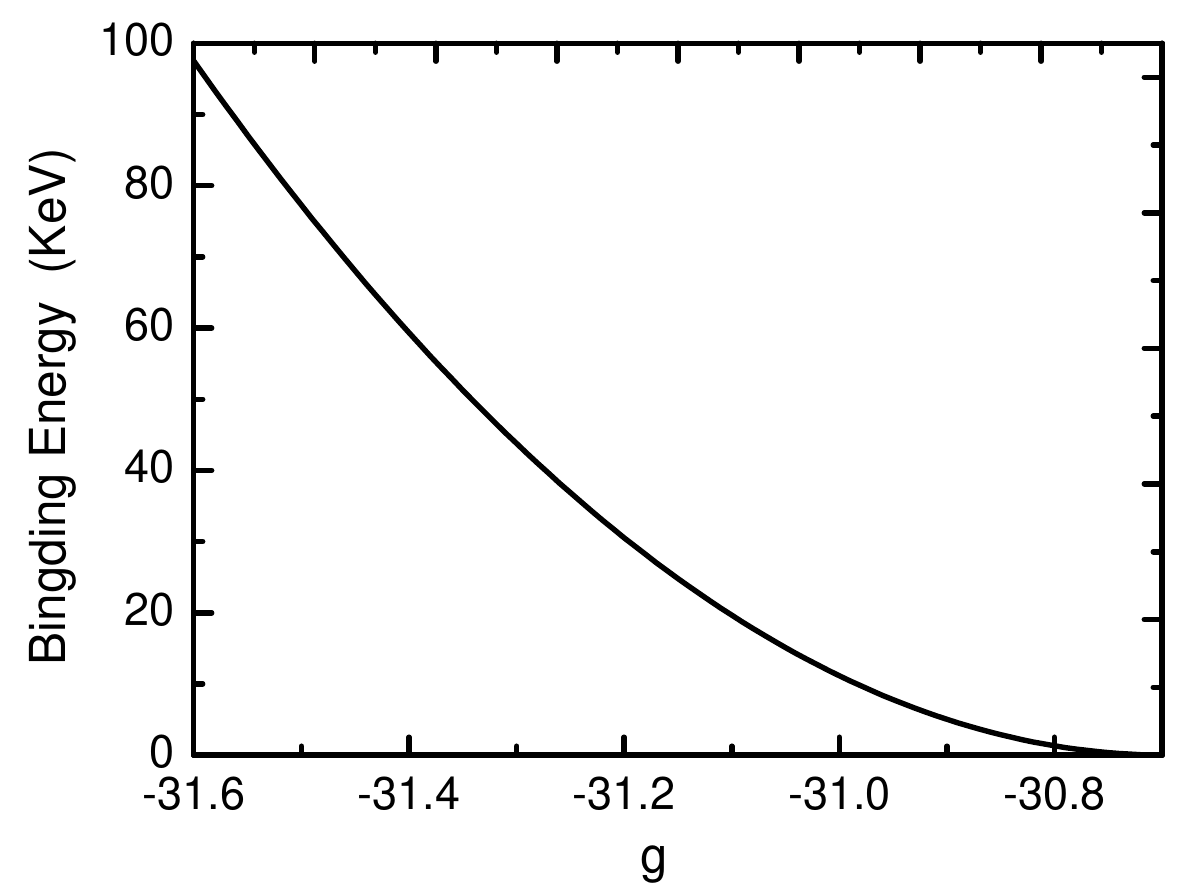}~~
\includegraphics[width=0.45\textwidth]{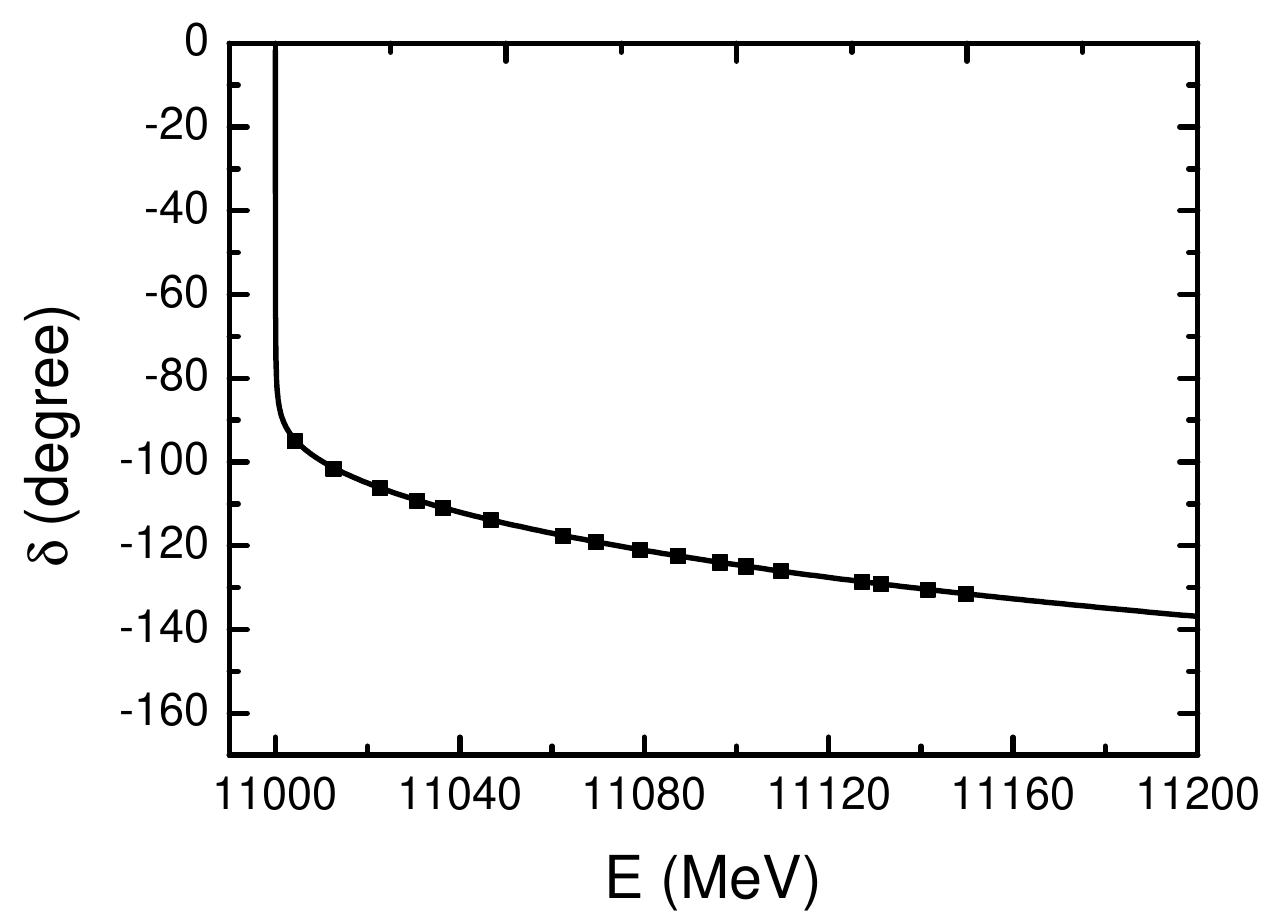}
\caption{Left panel: Binding energy of the $AN$ system as a function of the coupling
  constant $g$.
  Right panel: Phase shift for $AN\to AN$  scattering for $g= -31.0$. The black points are calculated
  from the corresponding energy levels depicted in Fig.~\ref{fig:ANtoANfinvol} using the L\"uscher equation.}
\label{fig:ANtoAN}
\end{figure}

First, we must fix the coupling constant $g$. In the left panel of Fig.~\ref{fig:ANtoAN},
we show the binding energy of the two-particle system as a function of the coupling $g$.
The latter is chosen in a range so that the binding is weak, and indeed at $g= -30.65$,
there is no more bound state. In what follows, we choose $g = -31.0$, for which one finds
a loosely bound state at $|E_B| = 11.15\,$keV. In the right panel of Fig.~\ref{fig:ANtoAN},
the corresponding scattering phase shift in the close-to-threshold region is shown, it
exhibits the typical features of a weakly bound state close to threshold.

\begin{figure}[t!]
\centering
\includegraphics[width=0.55\textwidth]{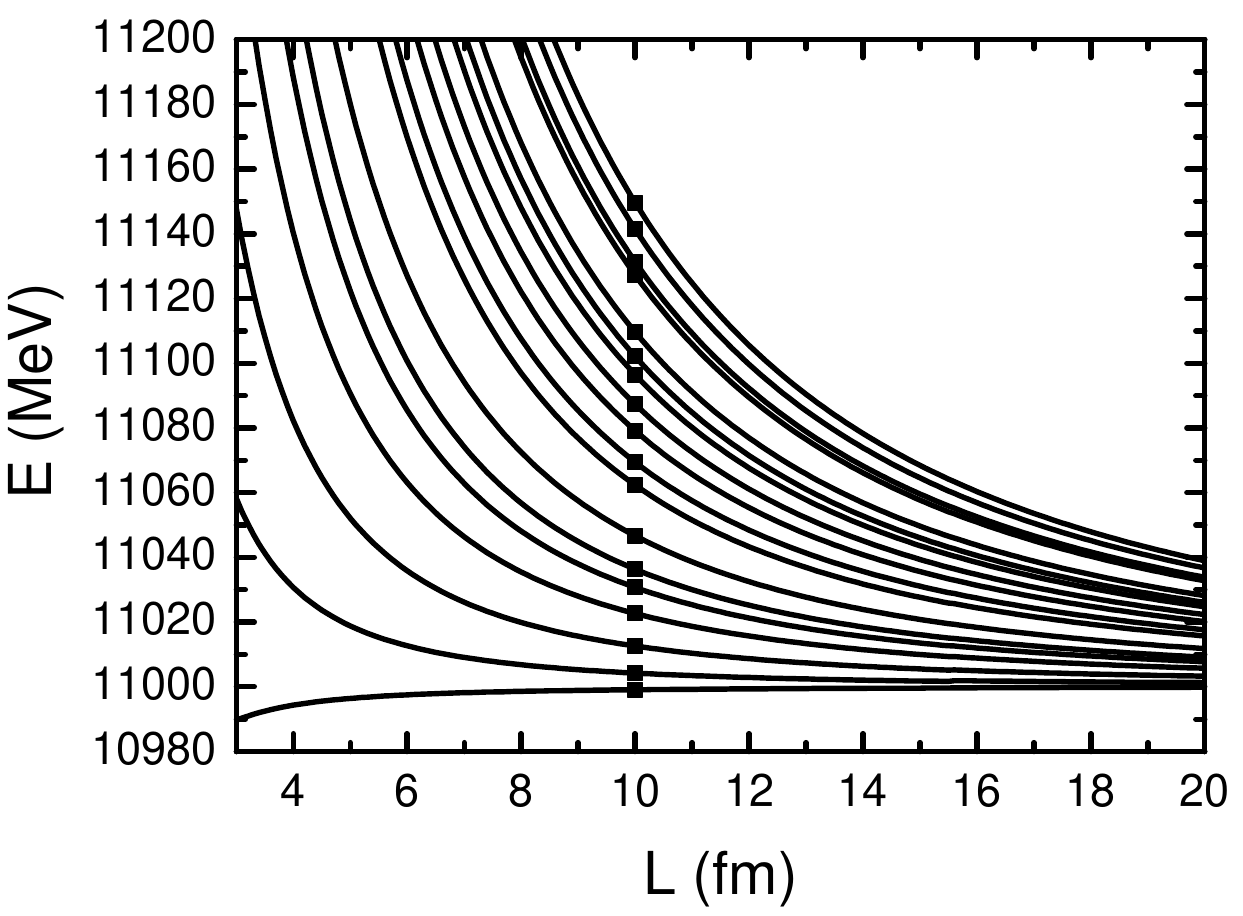}
\caption{Energy levels for the weakly bound $AN$ system in a finite volume $L^3$. The meaning of the
black squares at $L=10$~fm is explained in text, the also Fig.~\ref{fig:ANtoAN}.}
\label{fig:ANtoANfinvol}
\end{figure}

The corresponding energy levels in the finite volume are shown in  Fig.~\ref{fig:ANtoANfinvol}.
The bound state level is clearly visible, its bending downwards for smaller lattice sizes is
an expected finite volume effect. For sufficiently large $L$, these finite volume effects are visibly absent.

It is also instructive to compare our formalism with the L\"uscher equation \cite{Luscher:1986pf,Luscher:1990ux}.
For that, we pick out 17 energy levels at $L=10$ fm, and then we use following L\"uscher equation to calculate
the phase shifts from the corresponding  energy levels,
\begin{eqnarray}
\delta(q_E) &=&  \tan^{-1}\left(\frac{q_E L \sqrt{\pi}}{2  \mathcal{Z}(1; (\frac{q_EL}{2\pi})^2)} \right)+n\pi
\label{eq:Luscher}
\end{eqnarray}
where $q_E$ is on-shell momentum corresponding to the energy $E$,
\begin{eqnarray}
q_E&=&\frac{E}{2}
\sqrt{\left(1-\left(\frac{m_N+m_A}{E}\right)^2\right)\left(1-\left(\frac{m_N-m_A}{E}\right)^2\right)},
\end{eqnarray}
and $\mathcal{Z}(1;q^2)$ is well known Zeta-function. After regularization it can be calculated as follows,
\begin{eqnarray}
{\cal Z}(1; q^2)&=&\sum_{\vec{n}\in \mathbb{Z}^3}\frac{1}{\vec{n}^2-q^2}
\nonumber\\&=&
-\frac{1}{q^2}-8.91363292+16.53231596 q^2 
+ \sum_{\vec{n}\in \mathbb{Z}^3}\frac{q^4}{\vec{n}^4\left(\vec{n}^2-q^2\right)}~.
\end{eqnarray}
We find that the so calculated phase shifts are all on the phase shift curve  calculated
directly from the scattering function, see Fig.~\ref{fig:ANtoAN}.
This  shows that our calculation is consistent with the L\"uscher equation.

\section{Results for the full system}
\label{sec:3body}

\begin{figure}[t!]
\centering
\includegraphics[width=0.65\textwidth]{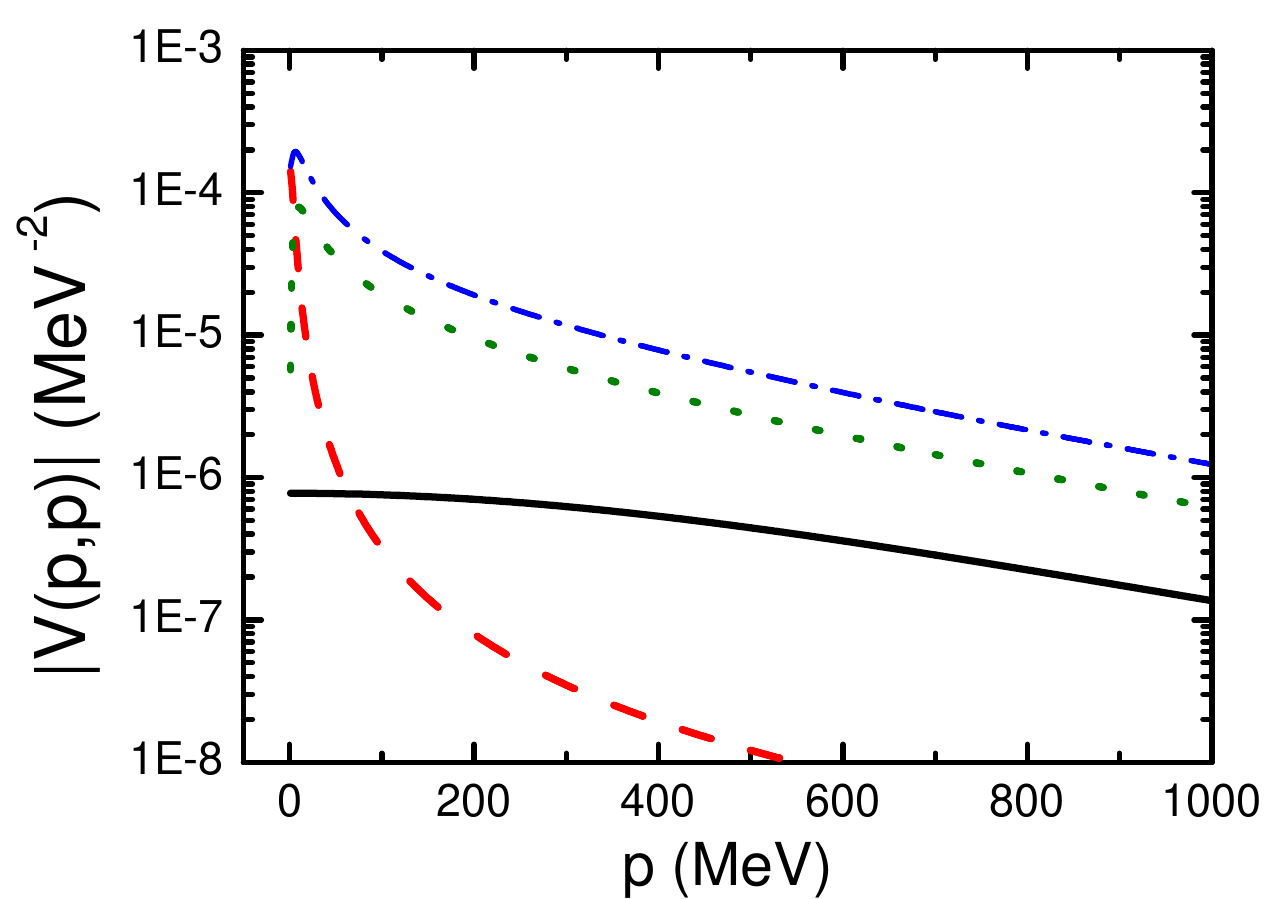}
\caption{Potentials in the $AN$ (black solid line) and the $BN$ system. In the latter case,
  two choices for the coupling $g_{NN}$ are made as discussed in the text (blue dash-dotted and
  green dotted lines). The red dashed line shows the attractive $BN$ potential from $A$-exchange.}
\label{fig:pot}
\end{figure}

Before showing the results, a few remarks are in order. We note that the atractive
potential of $BN\to BN$ by $A$-exchange  will have a sizeable magnitude  at  threshold,
because the propagator of the $A$-particle will be very close to zero since $B$ is a loose bound
state of $AN$. Similarly, the
repulsive potential $BN\to  BN$ generated by the triangle-loop will also have a
large value close to the threshold, since at that time, both nucleons can be to their mass shell.
We will therefore consider various choices to adjust the coupling $g_{NN}$, cf. Eq.~(\ref{eq:VNN}).
One is that these two contributions cancel exactly at threshold (case 1) and the other corresponds
to the case that the total
potential is repulsive (case 2).
In Fig.~\ref{fig:pot} we show the potential for various choices of the coupling $g_{NN}$.\\

\begin{figure}[t!]
\centering
\includegraphics[width=0.45\textwidth]{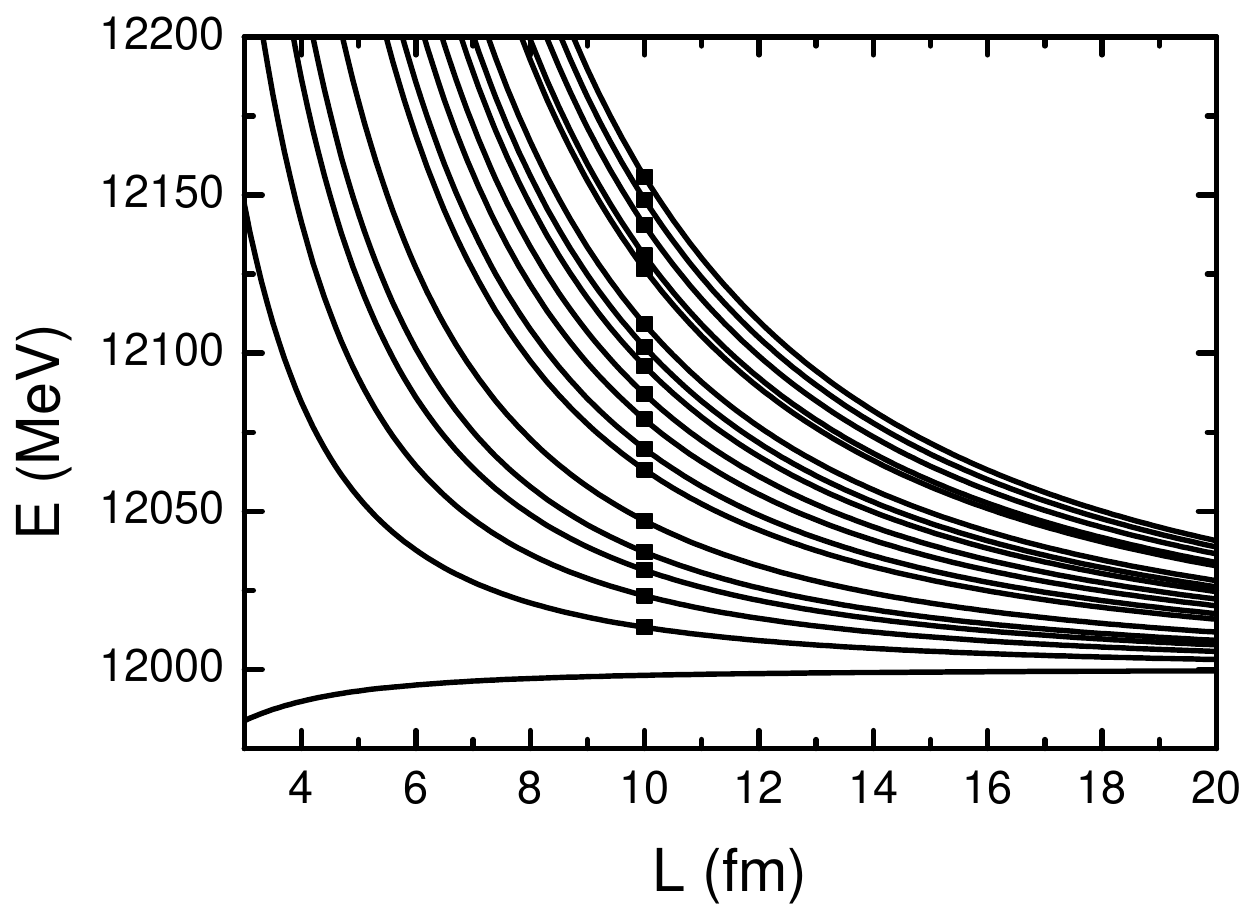}~~
\includegraphics[width=0.45\textwidth]{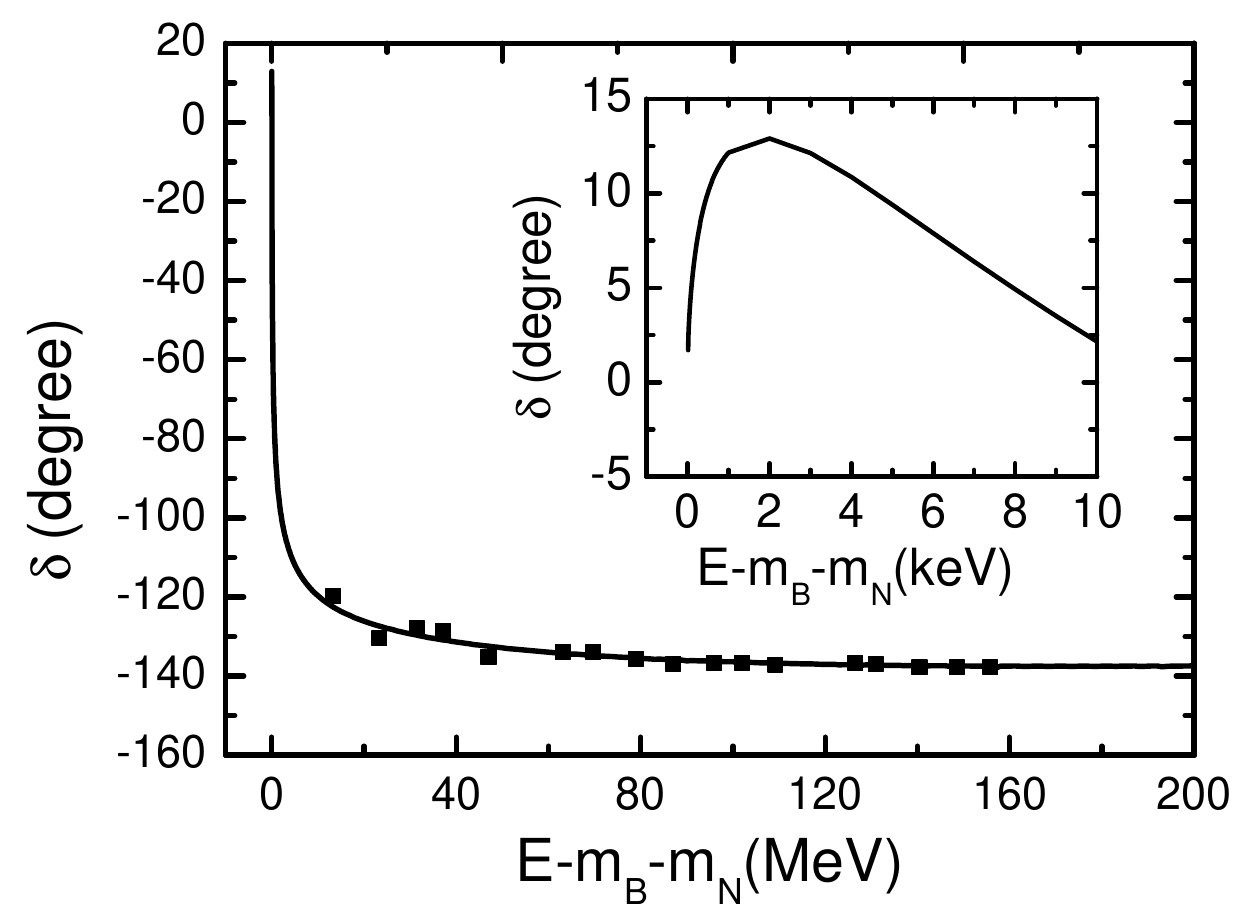}
\caption{Left panel: Energy levels for the  $BN$ system in a finite volume $L^3$ for $g_{NN} = 48.34$.
  Right panel: Phase shift for $BN\to BN$. The inset shows the phase shift very close to threshold. 
  The black points in the left panel are the data at $L=10$~fm. 
  The black points in the right panel are calculated from the corresponding data in the left panel with $L=10$~fm
  by using the L\"uscher equation as shown in Eq.~(\ref{eq:Luscher}). }
\label{fig:case1}
\end{figure}

\noindent
{\bf Case 1:}~With $g_{NN}=48.34$, there is a repulsive interaction between $BN\to BN$, but at threshold,
the potential is just zero. Above threshold the potential  increases fast and then drops almost
with the same slope as the potential from the loop.
Then from this potential, the corresponding finite volume spectrum can be computed as shown in
the left panel Fig.\ref{fig:case1}.
It is surprising that there is still a energy level below the threshold of $BN$ system since there
is pure repulsive potential.
This is due to the strange structure of potential at the threshold: At threshold, the potential is
exactly zero, and therefore the first term of full Hamiltonian matrix in the finite volume is just
the sum of the masses of $B$ and $N$.
On the other hand, in the finite volume the momentum is discrete, therefore, the off-diagonal
term in the Hamiltonian matrix will provide an attractive potential weather the original potential
is attractive or repulsive. 
Combing these two factors, the first energy level will be lower than the threshold in the
finite volume, especially for small lattice size.
The corresponding phase shift for $BN\to BN$ is shown in the right panel of Fig.~\ref{fig:case1}.
It is almost the same as that in $AN$ scattering, but we note that in the region very close to the
threshold the phase is increasing to about 10$^\circ$ as shown in the inset of right panel of
Fig.~\ref{fig:case1}.
The steep fall-off of the phase can be traced back to the fast decrease of the potential, as shown by the
green dotted curve in Fig.~\ref{fig:pot}.
We also check our method in comparison to the L\"uscher equation here.
The black points in the phase shift figure are calculated from the energy levels at $L=10$ fm which are
shown as black points in the left panel of Fig.\ref{fig:case1}.
Within small fluctuations, all of the points are consistent with the curves of phase shift which directly
calculated from scattering function.
These fluctuations will be discussed in the next case. 
\\

\begin{figure}[t!]
\centering
\includegraphics[width=0.45\textwidth]{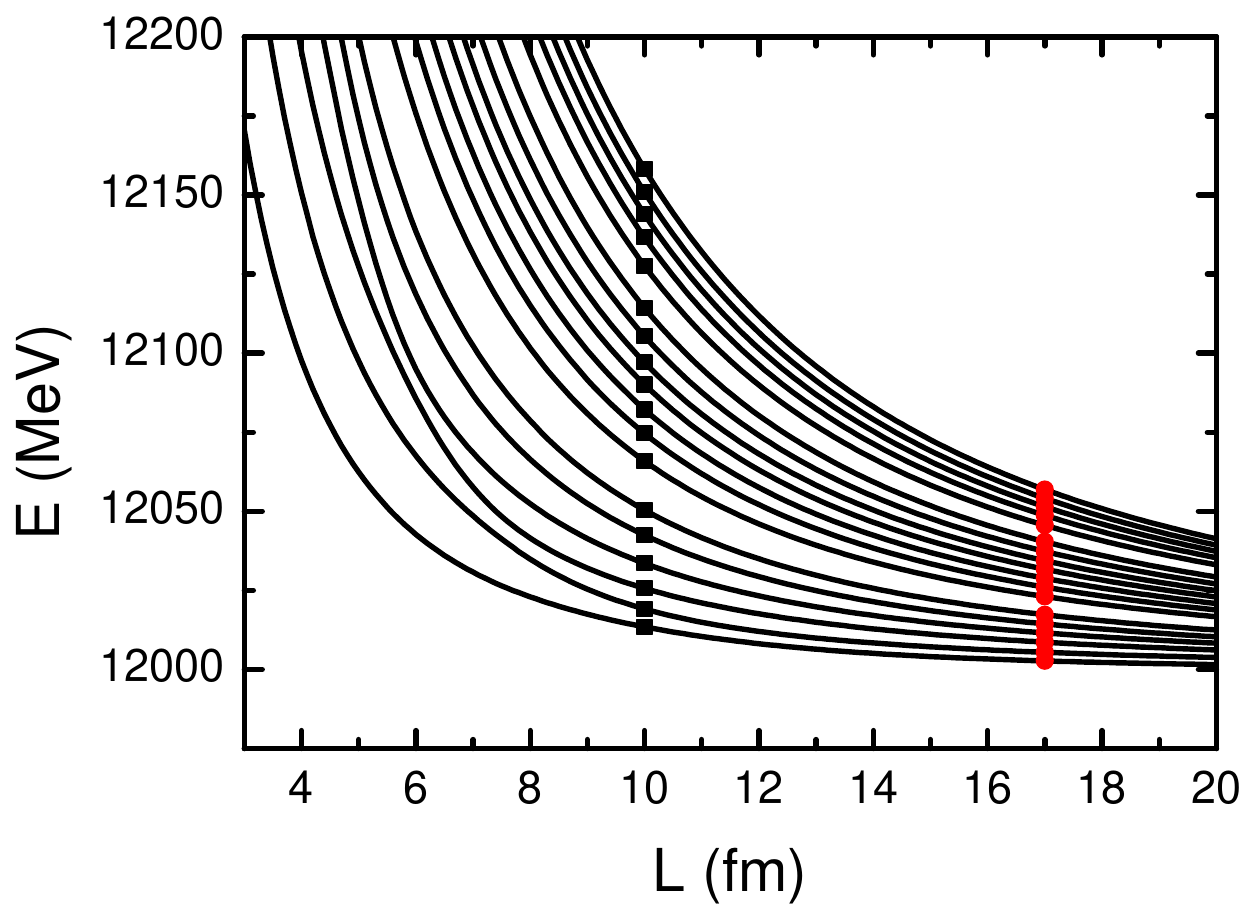}~~
\includegraphics[width=0.45\textwidth]{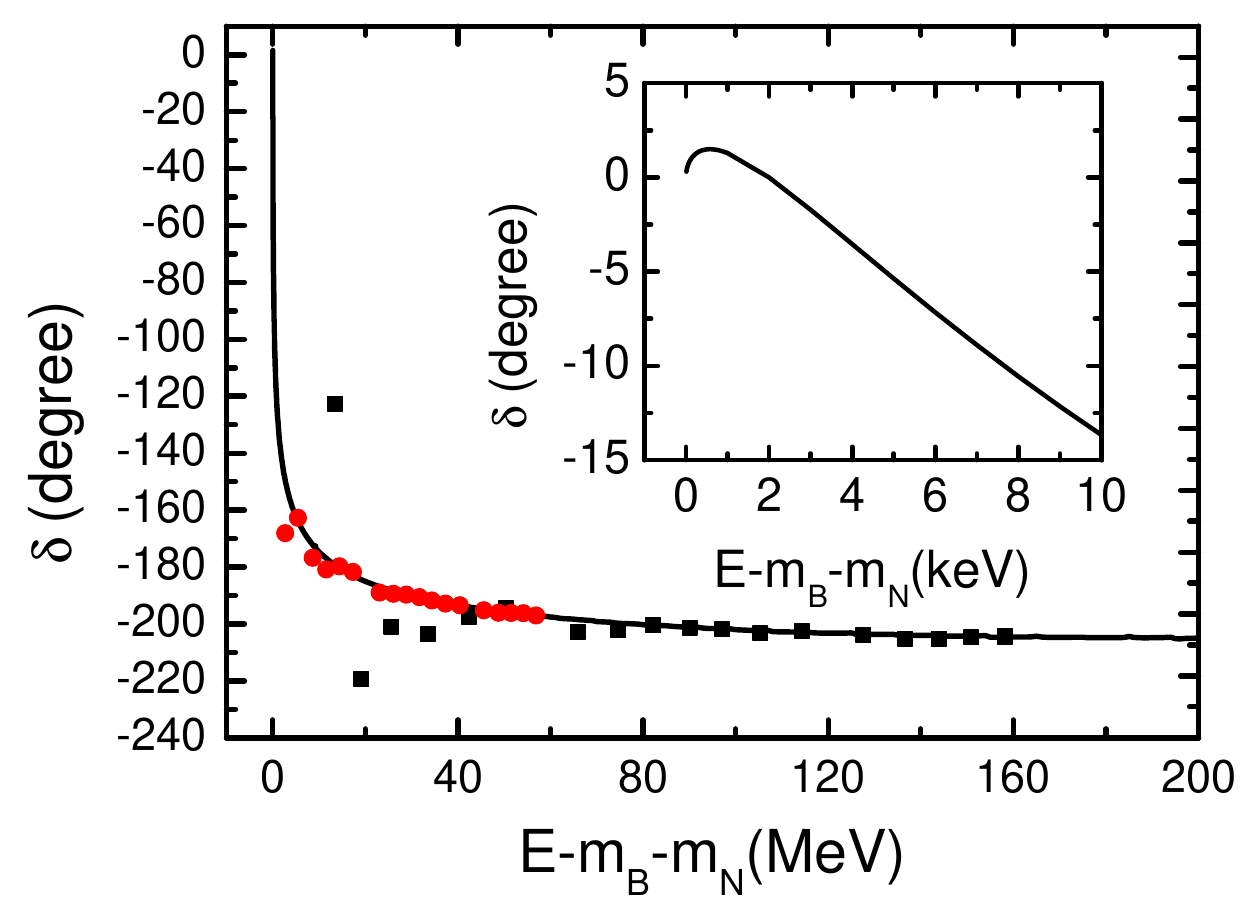}
\caption{Left panel: Energy levels for the  $BN$ system in a finite volume $L^3$ for $g_{NN} = 96.68$.
  Right panel: Phase shift for $BN\to BN$. The inset shows the phase shift very close to threshold.
  The black points and red circles in the left panel are the data at $L=10$ fm and $17$ fm, respectively.
  The black points and red circles in the right panel are calculated from the corresponding data in the left panel
  with $L=10$~fm and $17$~fm, respectively, using the L\"uscher equation as shown in Eq.~(\ref{eq:Luscher}).}
\label{fig:case2}
\end{figure}

\noindent
{\bf Case 2:}~With $g_{NN}=96.68$, there is a repulsive interaction and even at the threshold it has a
large value, although it still increases slightly with energy, see the blue dash-dotted line
in Fig.~\ref{fig:pot}. There is no bound state below the threshold in the finite volume spectrum
as shown in the left panel of Fig.~\ref{fig:case2}. 
In this case, the phase shift has a similar behaviour to that in the case~1 in the threshold region,
but the magnitude is much smaller, the largest phase shift here is around $1^\circ$. 
This corresponds to a potential barrier, so that the phase almost does not increase and very quickly
starts to fall as fast as the case~1,  shown as the blue dash-dotted curve in Fig.~\ref{fig:pot}.
Analogous to case1~1, we also check the consistency with L\"uscher equation. 
From the left panel of Fig.~\ref{fig:case2}, the first four points are far away from the curve of
the phase shift, which means that there is some inconsistency at the low energy levels. 
Actually, our methods has a systematic difference with the L\"uscher equation, which is the difference
between summation and interaction of a regular function as shown in the appendix of Ref.~\cite{Wu:2014vma}. 
This difference would be large when the regular function has some sharp structure and it is proportional to
$\exp(-Lm)$, where $m$ is the scale corresponding to the variation of the potential close to threshold.
In our case, the potential contributes significantly to the regular function and is very quickly
varying around the
threshold, therefore, such difference between summation and integration will be very large in this case.
However, in the $A+N \to A+N$ case, the potential is much more flat, and this leads  perfect consistency
between our method and the L\"uscher equation as shown in the left panel of Fig.~\ref{fig:ANtoAN}.
In other word, the fine structure at the threshold will be missing in the finite volume, when a too
small volume is used.
It can be resolved by increasing the lattice size, as shown in Fig.~\ref{fig:case2}.
The red circles are for the larger volume $L=17$ fm and are well consistent with the phase curve.
Therefore, in principle, our method is consistent with L\"ucher equation. 

From these observation we speculate that refined calculations will make it possible to
find a compact formula for the influence of the continuum on a weakly bound state on the lattice.

\section{Summary and outlook}
\label{sec:sum}

In this letter, we have made a first step to evaluate the influence of the continuum on weakly
bound states. We have shown that there is a visible interplay between the weak bound state $B$
in the two-particle system and the third particle, which leaves its traces in the lattice
energy spectrum. To draw more definite conclusions, the model used requires much refinement.
In a first step, the full three-body $ANN$ system should be investigated. Since the threholds of BN and 
ANN are very close, we  expect that the inelastic 
effects due to  breakup reaction $B + N \to A + N + N$ will affect the spectrum.
Then, the interaction
potentials have to be refined so that they resemble more closely the nuclear case. Also, higher
partial waves need to be included. Work along these lines in under way.

\section*{Acknowledgments}

We thank Dean Lee for a useful communication.
We acknowledge partial financial support from the Deutsche Forschungsgemeinschaft
(SFB/TRR~110, ``Symmetries and the Emergence of Structure in QCD'', grant no. TRR~110), by
the Chinese  Academy of Sciences (CAS) President's International Fellowship Initiative (PIFI)
(grant no. 2018DM0034) , by VolkswagenStiftung (grant no. 93562) and by the Fundamental Research
 Funds for the Central Universities. 

\section*{Appendix: Normalization of the scattering equation}
\label{app:1}

Here, we briefly discuss the normalization of our scattering T-matrix.
This normalization is similar to the formalism used in Ref.~\cite{Wu:2012md}.
Consider the s-wave of the process $AN\to AN$, with the scattering function given by
\begin{eqnarray}
&&T_H(E,|\vec{k}|, |\vec{k}'|) = 
V_H(|\vec{k}|, |\vec{k}'|) \nonumber\\
&&+ \int q^2 dq V_H(|\vec{k}|,q)
\frac{1}{E-\omega_A(q)-\omega_N(q)+i\epsilon}
T_H(E, q, |\vec{k}'|)~,\label{eq:Hscattering}
\end{eqnarray}
where $\omega_i(q)=\sqrt{m_i^2+q^2}$. 
Correspondingly, in the Bethe-Salpter  function,  where $k,k'$ are  four-momenta, takes the form
\begin{eqnarray}
T_L(P, k, k')&=&  V_L(P, k, k') \nonumber\\
&+&  \int d^4q V_L(P,k,q)
\frac{1}{q^2-m_A^2+i\epsilon}
\frac{1}{(P-q)^2-m_N^2+i\epsilon}
T_L(P, q, k')~.\nonumber\\ \label{eq:BSscattering}
\end{eqnarray}
Actually, Eq.~(\ref{eq:Hscattering}) can be recognized as the three-dimensional reduction of
Eq.~(\ref{eq:BSscattering}) by using
\begin{eqnarray}
&&\int d^4q 
\frac{1}{q^2-m_A^2+i\epsilon}
\frac{1}{(P-q)^2-m_N^2+i\epsilon} \nonumber\\
&\sim&
 \int q^2 dq 
\frac{1}{E-\omega_A(q)-\omega_N(q)+i\epsilon}\frac{1}{2\omega_A(q)2\omega_N(q)}~.
\label{eq:relation}
\end{eqnarray}
Therefore, we can get the relationship between the $V_H$ and $V_L$, 
\begin{eqnarray}
V_H(E,|\vec{k}|,|\vec{k}'|)
=\frac{2\pi}{\sqrt{2\omega_A(k)2\omega_N(k)}}
\int d \cos\theta V_L(P, k, k')
\frac{1}{\sqrt{2\omega_A(k')2\omega_N(k')}}~.
\label{eq:relation}
\end{eqnarray}
Note that we have been cavalier with some factors, such as $(2\pi)^n$, since these will be absorbed into
the coupling in $V_L$.
Also, this equation is the simple form of Eqs.(23-24) of Ref.~\cite{Wu:2012md}.


\begin{thebibliography}{99}

\bibitem{Dobaczewski:2007jc}
J.~Dobaczewski, N.~Michel, W.~Nazarewicz, M.~Ploszajczak and J.~Rotureau,
Prog. Part. Nucl. Phys. \textbf{59} (2007) 432
[arXiv:nucl-th/0701047 [nucl-th]].

\bibitem{Michel:2008pt}
N.~Michel, W.~Nazarewicz, M.~Ploszajczak and T.~Vertse,
J. Phys. G \textbf{36} (2009) 013101
[arXiv:0810.2728 [nucl-th]].

\bibitem{Meng:2015hta}
J.~Meng and S.~Zhou,
J. Phys. G \textbf{42} (2015) 093101
[arXiv:1507.01079 [nucl-th]].

\bibitem{Berggren:1968zz}
T.~Berggren,
Nucl. Phys. A \textbf{109} (1968) 265.

\bibitem{Berggren:1993zz}
T.~Berggren and P.~Lind,
Phys. Rev. C \textbf{47} (1993) 768.

\bibitem{Grasso:2001hf}
M.~Grasso, N.~Sandulescu, N.~Van Giai and R.~Liotta,
Phys. Rev. C \textbf{64} (2001) 064321.

\bibitem{Papadimitriou:2013ix}
G.~Papadimitriou, J.~Rotureau, N.~Michel, M.~Ploszajczak and B.~Barrett,
Phys. Rev. C \textbf{88} (2013)  044318
[arXiv:1301.7140 [nucl-th]].

\bibitem{Sun:2017unq}
Z.~Sun, Q.~Wu, Z.~Zhao, B.~Hu, S.~Dai and F.~Xu,
Phys. Lett. B \textbf{769} (2017) 227.
  
%
\bibitem{Baroni:2013fe}
S.~Baroni, P.~Navr\'atil and S.~Quaglioni,
Phys. Rev. C \textbf{87} (2013) 034326
[arXiv:1301.3450 [nucl-th]].

\bibitem{Romero-Redondo:2014fya}
C.~Romero-Redondo, S.~Quaglioni, P.~Navr\'atil and G.~Hupin,
Phys. Rev. Lett. \textbf{113} (2014) 032503
[arXiv:1404.1960 [nucl-th]].

\bibitem{Vorabbi:2019imi}
M.~Vorabbi, P.~Navr\'atil, S.~Quaglioni and G.~Hupin,
Phys. Rev. C \textbf{100} (2019) no.2, 024304
[arXiv:1906.09258 [nucl-th]].


\bibitem{Lee:2008fa}
D.~Lee,
Prog. Part. Nucl. Phys. \textbf{63} (2009) 117
[arXiv:0804.3501 [nucl-th]].

\bibitem{Lahde:2019npb}
T.~A.~Lähde and U.-G.~Mei{\ss}ner,
Lect. Notes Phys. \textbf{957} (2019) 1.


\bibitem{Epelbaum:2011md}
E.~Epelbaum, H.~Krebs, D.~Lee and U.-G.~Mei{\ss}ner,
Phys. Rev. Lett. \textbf{106} (2011) 192501
[arXiv:1101.2547 [nucl-th]].

\bibitem{Elhatisari:2015iga}
S.~Elhatisari, D.~Lee, G.~Rupak, E.~Epelbaum, H.~Krebs, T.~A.~L\"ahde, T.~Luu and U.-G.~Mei{\ss}ner,
Nature \textbf{528} (2015) 111
[arXiv:1506.03513 [nucl-th]].

\bibitem{Luscher:1986pf}
M.~L\"uscher,
Commun. Math. Phys. \textbf{105} (1986) 153.

  
\bibitem{Luscher:1990ux} 
  M.~L\"uscher,
  Nucl.\ Phys.\ B {\bf 354} (1991) 531.


\bibitem{Hall:2013qba} 
  J.~M.~M.~Hall, A.~C.-P.~Hsu, D.~B.~Leinweber, A.~W.~Thomas and R.~D.~Young,
  Phys.\ Rev.\ D {\bf 87} (2013) 094510 
  [arXiv:1303.4157 [hep-lat]].


\bibitem{Wu:2014vma}
J.~J.~Wu, T.~S.~Lee, A.~Thomas and R.~Young,
Phys. Rev. C \textbf{90} (2014)  055206
[arXiv:1402.4868 [hep-lat]].


\bibitem{Liu:2015ktc} 
  Z.~W.~Liu, W.~Kamleh, D.~B.~Leinweber, F.~M.~Stokes, A.~W.~Thomas and J.~J.~Wu,
  Phys.\ Rev.\ Lett.\  {\bf 116} (2016) 082004
  [arXiv:1512.00140 [hep-lat]].

\bibitem{Wu:2016ixr} 
  J.~J.~Wu, H.~Kamano, T.-S.~H.~Lee, D.~B.~Leinweber and A.~W.~Thomas,
  Phys.\ Rev.\ D {\bf 95} (2017) 114507
  [arXiv:1611.05970 [hep-lat]].


\bibitem{Liu:2016wxq} 
  Z.~W.~Liu, J.~M.~M.~Hall, D.~B.~Leinweber, A.~W.~Thomas and J.~J.~Wu,
  Phys.\ Rev.\ D {\bf 95} (2017) 014506
  [arXiv:1607.05856 [nucl-th]].

\bibitem{Wu:2017qve} 
  J.~j.~Wu, D.~B.~Leinweber, Z.~w.~Liu and A.~W.~Thomas,
  Phys.\ Rev.\ D {\bf 97} (2018) 094509
  [arXiv:1703.10715 [nucl-th]].


\bibitem{Li:2019qvh} 
  Y.~Li, J.~J.~Wu, C.~D.~Abell, D.~B.~Leinweber and A.~W.~Thomas,
  Phys.\ Rev.\ D {\bf 101} (2018) 114501
  [arXiv:1910.04973 [hep-lat]].


\bibitem{Kaplan:1996nv}
D.~B.~Kaplan,
Nucl. Phys. B \textbf{494} (1997) 471
[arXiv:nucl-th/9610052 [nucl-th]].

\bibitem{Kamalov:2000iy}
S.~S.~Kamalov, E.~Oset and A.~Ramos,
Nucl. Phys. A \textbf{690} (2001) 494
[arXiv:nucl-th/0010054 [nucl-th]].

\bibitem{Zhang:2019ykd}
X.~Zhang and J.~J.~Xie,
Chin. Phys. C \textbf{44} (2020)  054104
[arXiv:1906.07340 [nucl-th]].


\bibitem{Wu:2012md} 
  J.~J.~Wu, T.-S.~H.~Lee and B.~S.~Zou,
  Phys.\ Rev.\ C {\bf 85} (2012) 044002
  [arXiv:1202.1036 [nucl-th]].


\end{thebibliography}
\end{document}